\documentclass[12pt]{article}

\usepackage{amssymb,latexsym,amsmath, amscd}  

\newtheorem{theorem}{Theorem}[section]       % Numbered within each section
\newtheorem{proposition}[theorem]{Proposition}
\newtheorem{corollary}[theorem]{Corollary}

\newtheorem{definition}[theorem]{Definition}
\newtheorem{notation}[theorem]{Notation} 
\def\ln{\langle}\def\rn{\rangle}
\def\?#1?{{\bf #1}}
\def\.#1.#2.{{#1^{#2}}}
\def\kck #1.#2.{ {#1\choose #2} }
\def\tfo #1.#2.#3.#4.{{}_2 F_1\left({#1,#2\atop #3}\biggm| #4 \right)}
\def\xjm{\xi_j-\mu}
\def\pl{polynomial}
\def\ov #1.#2.{{#1\over #2}}
\def\e.#1.{e^{#1}}

\def\tfb #1.#2.#3.#4.{F_B\left({#1,#2\atop #3}\biggm| #4 \right)}

\def\KM{K}
\def\FFF#1#2{K^{(#1)}_{#2}}
\def\dir#1.{x^{N-{#1}}y^{#1}}
\def\dirs#1.{x^{N-{(#1)}}y^{#1}}  % dirac symmetric bracket

\newenvironment{proof}{{\sl Proof:}\quad}{\hfill{\qed}\\ \noindent}

\def\qed{\qquad\framebox[7pt]\medskip\noindent}

\numberwithin{equation}{section}

\title{Krawtchouk polynomials and Krawtchouk matrices}

\author{Philip Feinsilver and Jerzy Kocik\\         %}
\em Department of Mathematics\\
\em Southern Illinois University\\
\em Carbondale, IL 62901}
\date{}
\begin{document}

\maketitle

%\email{jkocik@siu.edu}

%\subjclass{
%05B20, %hadamard
%60G50, %random walk
%47A80, %tensor, operators 
%81P99, %qm computing
%46L53. %noncom probab
%}

%\keywords{Krawtchouk polynomials, Hadamard matrices, symmetric tensors, Krawtchouk encyclopedia}

\begin{abstract}
Krawtchouk matrices have as entries values of the Krawtchouk polynomials for nonnegative integer arguments.
We show how they arise as condensed Sylvester-Hadamard matrices via a binary shuffling function.
The underlying symmetric tensor algebra is then presented.\\

To advertise the breadth and depth of the field of Krawtchouk polynomials/matrices
through connections with various parts of mathematics, 
some topics  that are being developed into a Krawtchouk Encyclopedia are listed in the concluding section.
Interested folks are encouraged to visit the website 

\begin{center}
        {\tt http://chanoir.math.siu.edu/wiki/KravchukEncyclopedia}\\
\end{center}
\noindent
which is currently in a state of development. 
\end{abstract}

%%%%%%%%%%%%% What are Krawtchouk matrices
%===================  1   ==============================
\section{What are Krawtchouk matrices}
\label{sec:intro}

Of Sylvester-Hadamard matrices and Krawtchouk matrices,
the latter are less familiar, hence we start with them.

%----------------------
\begin{definition}\rm
The {\sl $N^{\mathrm{th}}$-order Krawtchouk  matrix $\FFF N{}$} is an
$(N+1)\times(N+1)$ matrix,
the entries of which are determined by the expansion:
\begin{equation}\label{eq:genkraw}
    (1+v)^{N-j} \; (1-v)^j = \sum_{i=0}^{N} \ v^i \KM^{(N)}_{ij}
\end{equation}
Thus, the polynomial 
$
              G(v)= (1+v)^{N-j}\; (1-v)^j
$
is the {\it generating function} for the 
row entries of the $j^{\mathrm{th}}$ column of $\KM^{(N)}$.
Expanding gives the explicit values of the matrix entries:
$$
\KM^{(N)}_{ij}= \sum_{k} (-1)^k {j \choose k}
      {N-j \choose i-k} .
$$ 
where  matrix indices run from $0$ to $N$.
\end{definition}

Here are the Krawtchouk  matrices of order zero, one, and two:
\begin{equation*}
\KM^{(0)}=\left[\begin{array}{rr}{ 1 }\end{array}\right] \qquad
\KM^{(1)}=\left[        
  \begin{array}{rr}
             1 &  1 \cr
             1 & -1 \cr 
             \end{array}\right]
 \qquad
\KM^{(2)}=\left[        
  \begin{array}{rrr}
             1 &  1 &  1 \cr
             2 &  0 & -2 \cr 
             1 & -1 &  1 \cr                 
             \end{array}\right]
 \end{equation*}
The reader is invited to see more examples in Table 1 of the Appendix. 
\\

The columns of Krawtchouk matrices may be considered 
\emph{generalized binomial coefficients}.
The rows define Krawtchouk \emph{polynomials}: for fixed order 
$N$, the $i^{\rm th}$ \emph{Krawtchouk polynomial} 
takes its corresponding values from the $i^{\rm th}$ row: 
\begin{equation} \label{eq:poly}
      k_i(j,N) = \KM^{(N)}_{ij} 
\end{equation}
 One can easily show that $k_i(j,N)$ can be given as a
polynomial of degree $i$ in the variable $j$. 
For fixed $N$, one has a system of $N+1$ polynomials orthogonal 
with respect to the symmetric binomial distribution.
\\

A  fundamental fact is that  
the square of a Krawtchouk  matrix is proportional to the identity matrix.
$$
(\FFF N{})^2 = 2^N \cdot I
$$
This property allows one to define 
a Fourier-like {\em Krawtchouk transform} on integer vectors. 
For more properties we refer the reader to \cite{FK}. In the present article, we focus
on Krawtchouk matrices as they arise from corresponding Sylvester-Hadamard matrices.
More structure is revealed through consideration of symmetric tensor algebra.

\bigskip
\noindent
{\bf Symmetric Krawtchouk matrices. }
When each column of a Krawtchouk matrix is multiplied by the corresponding binomial
coefficient, the matrix becomes symmetric.
In other words, define the  {\bf symmetric Krawtchouk  matrix} as
$$
S^{(N)}=\FFF N{}B^{(N)}
$$
where $B^{(N)}$ denotes the $(N+1)\times (N+1)$ diagonal matrix 
with binomial coefficients, 
$
          B^{(N)}_{ii}={N\choose i}
$, as its non-zero entries.

\bigskip\noindent
{\bf Example. }  For $N=3$, we have
$$
S^{(3)} =
       \left[\begin{array}{rrrr} 
                      1 &  1 &  1  &  1 \cr
                      3 &  1 & -1  & -3 \cr
                      3 & -1 & -1  &  3 \cr
                      1 & -1 &  1  & -1 \cr \end{array}\right]
       \left[\begin{array}{rrrr} 
                      1 &  0 & 0 & 0 \cr
                      0 &  3 & 0 & 0 \cr
                      0 &  0 & 3 & 0 \cr
                      0 &  0 & 0 & 1 \cr \end{array}\right]
=
      \left[\begin{array}{rrrr} 
                     1 &  3 &  3  &  1 \cr
                     3 &  3 & -3  & -3 \cr
                     3 & -3 & -3  &  3 \cr
                     1 & -3 &  3  & -1 \cr \end{array}\right] 
$$ 
\\

Some symmetric Krawtchouk  matrices are displayed in Table 2 of the
Appendix.  A study of the spectral properties of the symmetric Krawtchouk matrices
was initiated in work with Fitzgerald \cite{FF}.\bigskip

%%%%%%%%%%%%% Background note %%%%%%%%%%%%%
\noindent
{\bf Background note. }
Krawtchouk's polynomials were introduced  
by Mikhail Krawtchouk in the late 20's \cite{Kra, Kra2}. %1929,33
The idea of setting them in a matrix form appeared in the 1985 work 
of N. Bose \cite{B} %1985
on digital filtering in the context of the Cayley
transform on the complex plane. 
For some further development of this idea, see \cite{FK}.
\\

The Krawtchouk polynomials play an important r\^ole in many 
areas of mathematics.  Here are some examples: 
\bigskip
\begin{itemize}
\item
{\bf Harmonic analysis. }
As orthogonal polynomials, they appear in the classic work by
Sz\"ego \cite{Sze}. %1959
They have been studied from the point of view of 
harmonic analysis and special functions, 
e.g., in work of Dunkl \cite{Dun, DR}. %1974,76
Krawtchouk polynomials may be viewed as the discrete version of 
Hermite polynomials (see, e.g., \cite{AW}). 
\\
\item
{\bf Statistics. } Among the statistics literature we note particularly Eagleson \cite{Eag} %1969 
and  Vere-Jones \cite{V-J}. %1971
\\
\item
{\bf Combinatorics and coding theory. } Krawtchouk polynomials are essential in
MacWilliams' theorem on weight enumerators \cite{Lev,MS}, %1975, 97
and are a fundamental example in association schemes \cite{Del1,Del2,Del3}.   %1972,73
\bigskip
\item
{\bf Probability theory. }
In the context of the classical symmetric random walk,
it is recognized that Krawtchouk's polynomials are elementary 
symmetric functions in variables taking values $\pm1$. 
It turns out that the generating function (\ref{eq:genkraw}) is a
martingale in the parameter $N$ \cite{FS}. %1991
\bigskip
\item
{\bf Quantum theory. } 
Krawtchouk matrices interpreted as operators give rise  
to two new interpretations in the context of
both classical and quantum random walks  \cite{FK}.
The significance of the latter interpretation lies at the basis of quantum computing. 
\end{itemize}

\bigskip

Let us proceed to show the relationship between Krawtchouk matrices 
and Sylvester-Hadamard matrices.
\\
\\

%%%%%%%%%%%%% Krawtchouk matrices from Hadamard matrices %%%%%%%%%%%%%
%===================  3   ===========================================
\section{Krawtchouk matrices from Hadamard matrices}
\label{sec:result}

Taking the Kronecker (tensor) product of the initial matrix
\begin{equation*}
  H=\left[\begin{array}{cr} 
         1& 1\\
         1& -1 \end{array}\right]
\end{equation*}
with itself $N$ times defines the family of {\sl Sylvester-Hadamard matrices}.\\
  
(For a review of Hadamard matrices, see Yarlagadda and Hershey \cite{YH}.) %1997
%For statisticians, they point out that in Yates' factorial analysis, 
%the Hadamard transform provides a useful nonparametric test for association.
\bigskip

\begin{notation}
Denote the Sylvester-Hadamard matrices, tensor (Kronecker) powers
of the fundamental matrix $H$, by
\begin{displaymath}
  H^{(N)} = H^{\otimes N} =
      \underbrace{H\otimes H\otimes \cdots \otimes H}_{N\hbox{ times}} 
\end{displaymath}
\end{notation}
The first three Sylvester-Hadamard matrices are: 
\def\b{\circ}   \def\a{\bullet}
$$
H^{(1)}=\left[
\begin{array}{cc}
\a &\a  \cr
\a &\b  \cr
\end{array}\right]
\quad
H^{(2)}=\left[\begin{array}{cccc}
\a &\a &\a &\a  \cr
\a &\b &\a &\b   \cr 
\a &\a &\b &\b   \cr
\a &\b &\b &\a   \cr
\end{array}\right]
\quad
H^{(3)}=\left[\begin{array}{cccccccc}
\a &\a &\a &\a  &\a &\a &\a &\a    \cr
\a &\b &\a &\b  &\a &\b &\a &\b    \cr
\a &\a &\b &\b  &\a &\a &\b &\b    \cr
\a &\b &\b &\a  &\a &\b &\b &\a    \cr
\a &\a &\a &\a  &\b &\b &\b &\b    \cr 
\a &\b &\a &\b  &\b &\a &\b &\a    \cr 
\a &\a &\b &\b  &\b &\b &\a &\a    \cr 
\a &\b &\b &\a  &\b &\a &\a &\b    \cr 
\end{array}\right] 
$$ 
where, to emphasize the patterns, we use
$\bullet$ for 1 and $\circ$ for -1. See Table 3 of the Appendix for these matrices up to order 5.
\\

For $N=1$, the Hadamard matrix coincides with 
the Krawtchouk matrix:  $H^{(1)}=K^{(1)}$.
Now we wish to see how the two classes of matrices are related for higher $N$.
It turns out that appropriately contracting (condensing) Hadamard-Sylvester matrices 
yields corresponding symmetric Krawtchouk  matrices.\bigskip

The problem is that the tensor products disperse
the columns and rows that have to be 
summed up to do the contraction. 
We need to identify the right sets of indices.
\\
\begin{definition}\rm
Define the {\sl binary shuffling function} 
as the function
$$
         w\colon {\bf N} \to {\bf N}
$$
giving the ``binary weight" of an integer.
That is, let $n=\sum_k d_k 2^k$ be the binary expansion of the
number $n$. Then $w(n)=\sum_k d_k$, 
the number of ones in the representation. 
\end{definition}
Notice that, as sets,
$$       
       w(\{0,1,\ldots, 2^N-1\}) =    \{0, 1,\ldots, N\} 
$$
Here are the first 16 values of $w$ listed for the 
integers running from 0 through $2^4-1=15$:
$$
0 \quad 1 \quad 1 \quad 2 \quad 1 \quad 2 \quad 2 \quad 3 \quad
1 \quad 2 \quad 2 \quad 3 \quad 2 \quad 3 \quad 3 \quad 4 
$$
The shuffling function can be defined recursively. Set 
\hfill\break $w(0)=0$ and
\begin{equation}\label{eq:dynamic}
  w(2^N + k) = w(k)+1 
\end{equation}
for $0\leq k < 2^N$.
One can thus create the sequence of values of the shuffling function 
by starting with $0$ and then appending to the current string
of values a copy of itself with values increased by $1$:
\begin{displaymath}
0
\ \rightarrow\
01
\ \rightarrow\
0112
\ \rightarrow\
01121223
\ \rightarrow\ 
\hbox{\ldots} 
\end{displaymath}

Now we can state the result;

\begin{theorem}
Symmetric Krawtchouk matrices are reductions of Hadamard matrices
as follows:
$$
S^{(N)}_{ij} = \sum_{w(a)=i \atop w(b)=j} H^{(N)}_{ab}
$$
\end{theorem}
~
\\
\\
\noindent
{\bf Example. }
Let us see the transformation for $H^{(4)} \to S^{(4)}$
(recall that $\bullet$ stands for $1$, and $\circ$ for $-1$).
Applying the binary 
shuffling function to $H^{(4)}$, mark the rows and columns 
accordingly:

$$
\bordermatrix{
  &0  &1  &1  &2   &1  &2  &2  &3   &1  &2  &2  &3   &2  &3  &3  &4   \cr 
0 &\a &\a &\a &\a  &\a &\a &\a &\a  &\a &\a &\a &\a  &\a &\a &\a &\a  \cr
1 &\a &\b &\a &\b  &\a &\b &\a &\b  &\a &\b &\a &\b  &\a &\b &\a &\b  \cr 
1 &\a &\a &\b &\b  &\a &\a &\b &\b  &\a &\a &\b &\b  &\a &\a &\b &\b  \cr
2 &\a &\b &\b &\a  &\a &\b &\b &\a  &\a &\b &\b &\a  &\a &\b &\b &\a  \cr
1 &\a &\a &\a &\a  &\b &\b &\b &\b  &\a &\a &\a &\a  &\b &\b &\b &\b  \cr
2 &\a &\b &\a &\b  &\b &\a &\b &\a  &\a &\b &\a &\b  &\b &\a &\b &\a  \cr
2 &\a &\a &\b &\b  &\b &\b &\a &\a  &\a &\a &\b &\b  &\b &\b &\a &\a  \cr
3 &\a &\b &\b &\a  &\b &\a &\a &\b  &\a &\b &\b &\a  &\b &\a &\a &\b  \cr
1 &\a &\a &\a &\a  &\a &\a &\a &\a  &\b &\b &\b &\b  &\b &\b &\b &\b  \cr
2 &\a &\b &\a &\b  &\a &\b &\a &\b  &\b &\a &\b &\a  &\b &\a &\b &\a  \cr
2 &\a &\a &\b &\b  &\a &\a &\b &\b  &\b &\b &\a &\a  &\b &\b &\a &\a  \cr
3 &\a &\b &\b &\a  &\a &\b &\b &\a  &\b &\a &\a &\b  &\b &\a &\a &\b  \cr
2 &\a &\a &\a &\a  &\b &\b &\b &\b  &\b &\b &\b &\b  &\a &\a &\a &\a  \cr
3 &\a &\b &\a &\b  &\b &\a &\b &\a  &\b &\a &\b &\a  &\a &\b &\a &\b  \cr
3 &\a &\a &\b &\b  &\b &\b &\a &\a  &\b &\b &\a &\a  &\a &\a &\b &\b  \cr
4 &\a &\b &\b &\a  &\b &\a &\a &\b  &\b &\a &\a &\b  &\a &\b &\b &\a  \cr
}
$$
~
\\
The contraction is performed by summing columns with the same index,
then summing rows in similar fashion.
One checks from the given matrix that
indeed this procedure gives the symmetric Krawtchouk
matrix $S^{(4)}$:  
$$
S^{(4)}=
\bordermatrix{
   &\hfill0  &\hfill1  &\hfill2  &\hfill3   &\hfill4   \cr 
0  &\hfill1  &\hfill4  &\hfill6  &\hfill4   &\hfill1   \cr
1  &\hfill4  &\hfill8  &\hfill0  &\hfill-8  &\hfill-4  \cr
2  &\hfill6  &\hfill0  &\hfill-12 &\hfill0   &\hfill6   \cr
3  &\hfill4  &\hfill-8 &\hfill0  &\hfill8   &\hfill-4  \cr
4  &\hfill1  &\hfill-4 &\hfill6  &\hfill-4  &\hfill1  \cr
}
$$
~
\\

Now we give a method for transforming the $N^{\rm th}$ (symmetric) Krawtchouk matrix
into the $N+1^{\rm st}$.

\begin{definition}\rm
\label{d:shuffle}
The {\sl square contraction $r(M)$} of a $2n\times 2n$ matrix $M_{ab}$, \hfill\break
$1\le a,b\le 2n$, is the $(n+1)\times (n+1)$ matrix with entries
\begin{equation*}
(rM)_{ij} = \sum_{a=2i,\; 2i+1 \atop b=2j,\; 2j+1}
M_{ab}
\end{equation*}
$0\le i,j\le n$,  where the values of $M_{ab}$ with $a$ or $b$ outside of the range $(1,\ldots, 2n)$ 
are taken as zero. 
\end{definition}

\begin{theorem}
Symmetric Krawtchouk matrices satisfy:
$$
S^{(N+1)} = r ( S^{(N)}\otimes H )
$$
with $S^{(1)}=H$.
\end{theorem}

\noindent
{\bf Example. }  Start with symmetric Krawtchouk matrix of order 2:
$$
S^{(2)} = 
\left[ \begin{array}{rrr}   
                  1 & 2 & 1 \cr
                  2 & 0 & -2 \cr
                  1 & -2& 1  \cr\end{array} \right]
$$
Take the tensor product with $H$:
$$
S^{(2)}\otimes H = 
\left[ \begin{array}{rrrrrr}   
                  1&1 & 2&2 & 1&1 \cr
                  1&-1 & 2&-2 & 1&-1 \cr
                  2&2 & 0&0 & -2&-2 \cr
                  2&-2 & 0&0 & -2&2 \cr
                  1&1 & -2&-2& 1&1  \cr
                  1&-1 & -2&2& 1&-1  \cr\end{array} \right]
$$
surround with zeros and contract:                
$$
r (S^{(2)}\otimes H) = r
\left[ \begin{array}{rr|rr|rr|rr} 
                0& 0& 0& 0& 0&  0&   0 &0 \cr 
                0& 1& 1& 2& 2 & 1&   1 &0 \cr
                \multispan{8} \hrulefill\cr
                0& 1&-1&2&  -2 & 1& -1&0  \cr
                0& 2&2 &  0&0 & -2& -2 &0 \cr
                \multispan{8} \hrulefill\cr                
                0& 2&-2 & 0&0 & -2& 2 &0  \cr
                0& 1&1 & -2&-2& 1&  1  &0 \cr
                \multispan{8} \hrulefill\cr               
                0& 1&-1 &-2&2& 1&   -1 &0  \cr
                0& 0& 0&0&  0& 0&   0 &0 \cr \end{array}\right]
%r (S^{(2)}\otimes H) = r
%\left[ \matrix{ 0& 0&\vdots& 0& 0&\vdots& 0&  0&   \vdots&0 &0 \cr 
%                0& 1&\vdots& 1& 2&\vdots& 2 & 1&   \vdots&1 &0 \cr
%                \multispan{11} \hrulefill\cr
%                0& 1&\vdots&-1&2& \vdots& -2 & 1& \vdots&-1&0  \cr
%                0& 2&\vdots&2 &  0&\vdots&0 & -2& \vdots&-2 &0 \cr
%                \multispan{11} \hrulefill\cr                
%                0& 2&\vdots&-2 & 0&\vdots&0 & -2& \vdots&2 &0  \cr
%                0& 1&\vdots&1 & -2&\vdots&-2& 1&  \vdots&1  &0 \cr
%                \multispan{11} \hrulefill\cr               
%                0& 1&\vdots&-1 &-2&\vdots&2& 1&   \vdots&-1 &0  \cr
%                0& 0&\vdots& 0&0&  \vdots&0& 0&   \vdots&0 &0 \cr } \right]
%
=
\left[ \begin{array} {rrrr}
                1& 3&  3& 1 \cr 
                3& 3& -3& -3 \cr
                3& -3& -3& 3 \cr
                1&-3&  3& -1 \cr\end{array} \right]
$$
~
\\
\begin{corollary}
Krawtchouk matrices satisfy:
$$
\KM^{(N+1)} = r ( \KM^{(N)} B^{(N)} \otimes H ) (B^{(N+1)})^{-1}
$$
where $B$ is the diagonal binomial matrix.
\end{corollary}

Note that starting with the $2\times2$ identity matrix, $I$, set $I^{(1)}=I$, \\ $I^{(N+1)}=r(I^{(N)}\otimes I)$. 
Then, in fact, $I^{(N)}=B^{(N)}$.\bigskip

Next, we present the algebraic structure underlying these remarkable properties.

%%%%%%%%%%%%% Krawtchouk matrices and symmetric tensors %%%%%%%%%%%%%
%========================= 3 ======================

\section{Krawtchouk matrices and symmetric tensors}
\label{sec:quantum}
\def\symm{{\tt symm}}

Given a $d$-dimensional vector space $V$ over ${\bf R}$,
one may construct a $d^N$-dimensional space $V^{\otimes N}$ ,
the $N$-fold tensor product of $V$, and, as well, a 
$
     {d+N-1\choose N}
$-dimensional {\sl symmetric tensor space} $V^{\otimes_s N}$. 
There is a natural map 
$$
  \symm\colon V^{\otimes N} \ \longrightarrow \ V^{\otimes_S N} 
$$
which, for homogeneous tensors, is defined via
$$
\symm\,(v\otimes w\otimes \ldots )= \hbox{symmetrization of}\,(v\otimes w\otimes \ldots)
$$

For computational purposes, it is convenient to use the fact that the
symmetric tensor space of order $N$ of a $d$-dimensional vector space is isomorphic to the 
space of polynomials in $d$ variables homogeneous of degree $N$.\bigskip

Let $\{e_1, e_2, \ldots e_d\}$ be a basis of $V$. Map $e_i$ to $x_i$, replace tensor products by
multiplication of the variables, and extend by linearity.  For example,
$$
2e_1\otimes e_2 +3 e_2\otimes e_1 -7e_3\otimes e_2 \ \longrightarrow \
5x_1x_2 -7x_2x_3
$$
thus identifying basis (elementary) tensors in $V^{\otimes N}$ that are equivalent under any permutation.
\\

This  map induces a map on certain linear operators.
Suppose $A\in \hbox{End}(V)$ is a linear transformation on $V$.  This induces
a linear transformation $A_N=A^{\otimes N}\in \hbox{End}(V^{\otimes N}) $ defined on elementary tensors by:
$$
A_N(v\otimes w\otimes\ldots) = 
A(v)\otimes A(w)\otimes\ldots
$$
Similarly, a linear operator on the symmetric tensor spaces is induced so 
that the following diagram commutes:
$$
\begin{CD}
 V^{\otimes N}    @>A_N>>            V^{\otimes N}   \\
 @V{\symm}VV                          @V{\symm}VV    \\
 V^{\otimes_s N}  @>\overline A_N>>  V^{\otimes_s N}
\end{CD}
$$
~
\\
This can be understood by examining the action on polynomials.
We call  $\overline{A}_N$ the 
{\sl symmetric representation of $A$ in degree $N$}.  
Denote the matrix elements of $\overline{A}_N$ by $\overline{A}_{mn}$. 
If $A$ has matrix entries
$A_{ij}$, let
$$
       y_i=\sum_j A_{ij}\,x^j 
$$
It is convenient to label variables with indices from $0$ to $\delta=d-1$.
Then the matrix elements of the symmetric representation
are defined by the expansion:
$$
   y_0^{m_0}\cdots y_\delta^{m_\delta}=
               \sum_n \overline{A}_{mn}\,
                     x_0^{n_0}\cdots x_\delta^{n_\delta}
$$
with multi-indices $m$ and $n$ homogeneous of degree $N$.
\\

Mapping to the symmetric representation is an algebra homomorphism, i.e., 
$$
\overline{AB}=\overline{A}\,\overline{B}
$$
Explicitly, in matrix notation, 
$ 
\overline{(AB)}_{mn}=
          \sum\limits_r \,(\overline{A})_{mr}\,
                 (\overline{B})_{rn}  \;.
$ 
\\
\\

Now we are ready to state our result
\bigskip

\begin{proposition} 
\label{prop:h2k}
For each $N>0$, the symmetric representation of the $N^{\rm th}$ Sylvester-Hadamard 
matrix equals the transposed $N^{\rm th}$ Krawtchouk matrix:
$$ 
    (\overline{H}_N)_{ij} = \FFF{N}{ji}\;.
$$
\end{proposition}

\begin{proof}
Writing $(x,y)$ for $(x_0,x_1)$, we have in degree $N$ for the 
$k^{\mathrm{th}}$ component:
$$
        (x+y)^{N-k}(x-y)^k=\sum_l \overline{H}_{kl} \,\dir l.
$$
Substituting $x=1$ yields
the generating function (\ref{eq:genkraw}) for the Krawtchouk matrices with the
coefficient of $y^l$ equal to $\FFF{N}{lk}$. Thus the result.
\end{proof}
~
\\
\\

Insight  into these correspondences can be gained by splitting 
the fundamental Ha\-da\-mard matrix $H$ ($=K^{(1)}$) into two special symmetric $2\times2$ operators:
$$ 
         F=\left[\begin{array}{cc}0&1\cr 1&0\cr\end{array}\right],    \qquad
         G=\left[\begin{array}{cc}1&0\cr \hfill 0&-1\cr\end{array}\right]
$$
so that 
$$
H=F+G = \left[\begin{array}{cc} 1 & \hfill 1 \cr
                       1 & -1 \cr\end{array}\right]
$$
One can readily check that 
\begin{eqnarray}
       & F^2=G^2=I & \nonumber\\ \label{eq:H2}
       & FH =HG    \qquad\hbox{and}\qquad GH=HF&
\end{eqnarray}
The first of the second pair of equations may be viewed as 
the spectral decomposition of $F$ and we can interpret the Hadamard
matrix as diagonalizing $F$ into $G$. Taking transposes gives the second equation of
(\ref{eq:H2}).\\

Now we proceed to the interpretation leading to a symmetric Bernoulli quantum random walk (\cite{FK}).
For this interpretation, the Hilbert space of states is represented 
by the $N^{\rm th}$ tensor power of the original 2-dimensional space $V$,
that is, by the $2^N$-dimensional Hilbert space $V^{\otimes N}$\ .
Define the following linear operator on $V^{\otimes N}$:
\begin{eqnarray*}
X_F &=& F\otimes I \otimes\cdots\otimes I          \\
     &&+ I\otimes F\otimes I \otimes\cdots\otimes I \\
     && +\ldots                                    \\
     && +I\otimes I \otimes\cdots\otimes F \\
    &=& f_1+f_2+\ldots+f_i+\ldots+f_N
\end{eqnarray*}
each term describing a ``flip" at the $i^{\rm th}$ position (cf. \cite{H,Si}).
Analogously, we define:
\begin{eqnarray*}
X_G &=& G\otimes I \otimes\cdots\otimes I\\
    & & +I\otimes G\otimes I \otimes\cdots\otimes I\\
    & & +\ldots \\
    & & +I\otimes I \otimes\cdots\otimes G \\
    &=& g_1+g_2+\ldots+g_i+\ldots+g_N
\end{eqnarray*}

From equations (\ref{eq:H2}) we see that our $X$-operators intertwine
the Sylvester-Hadamard matrices:  
$$
            X_F H^{(N)} = H^{(N)} X_G \qquad\hbox{and}\qquad X_G H^{(N)} = H^{(N)} X_F
$$
Since products are preserved in the process of passing to
the symmetric tensor space, we get
\begin{equation} \label{eq:spec}
       \overline{X}_F\,\overline{H}_N  = \overline{H}_N  \,\overline{X}_G \qquad \hbox{and}\qquad
       \overline{X}_G\,\overline{H}_N = \overline{H}_N  \,\overline{X}_F 
\end{equation}
the bars indicating the corresponding induced maps.\bigskip

We have seen in Proposition \ref{prop:h2k} how to calculate $\overline{H}_N$ from the action of $H$ on
polynomials in degree $N$. For symmetric tensors we have the
components in degree $N$, namely 
$ 
      x^{N-k}y^k
$, 
for $0\le k\le N$, where for convenience we write $x$ for $x_0$ and $y$ for $x_1$.
Now consider the generating function for the elementary symmetric
functions in the quantum variables $f_j$. This is the $N$-fold
tensor power
$$ 
     {\mathcal F}_N (t) = (I+tF)^{\otimes N}
                    = I^{\otimes N}+t\,X_F+\cdots
$$
noting that the coefficient of $t$ is $X_F$. 
Similarly, define
$$
     {\mathcal G}_N(t) = (I+tG)^{\otimes N}
                   = I^{\otimes N}+tX_G+\cdots
$$
  From $(I+tF)H=H(I+tG)$ we have
$$
   {\mathcal F}_NH^{(N)}  =  H^{(N)}{\mathcal G}_N        
   \qquad\hbox{and}\qquad 
   \overline{\mathcal F}_N \,\overline{H}_N  
 = \overline{H}_N \,\overline{\mathcal G}_N
$$
The difficulty is to calculate the action on the symmetric tensors for
operators, such as $X_F$, that are not pure tensor powers. However,
from ${\mathcal F}_N(t)$ and ${\mathcal G}_N(t)$ we can 
recover $X_F$ and $X_G$ via
$$
          X_F=\frac{d}{dt}\biggm|_{t=0}(I+tF)^{\otimes N},
     \qquad
          X_G=\frac{d}{dt}\biggm|_{t=0}(I+tG)^{\otimes N}
$$
with corresponding relations for the barred operators.
Calculating on polynomials yields the desired results as follows.
$$ 
          I+tF = \left[\begin{array}{cc}1&t\cr t&1\cr\end{array}\right],\qquad
          I+tG = \left[\begin{array}{cc}1+t&0\cr 0&1-t\cr\end{array}\right]
$$
In degree $N$, using $x$ and $y$ as variables, we get the 
$k^{\mathrm{th}}$ component for $\overline{X}_F$ and $\overline{X}_G$
via
\begin{eqnarray*}
    \frac{d}{dt}\biggm|_{t=0}(x+ty)^{N-k}(tx+y)^k
                     &=& (N-k)\,\dirs k+1. +k\,\dirs k-1. 
                   \end{eqnarray*}
and since $I+tG$ is diagonal,
\begin{eqnarray*}
    \frac{d}{dt}\biggm|_{t=0}(1+t)^{N-k}(1-t)^k\,\dir k.
                     &=& (N-2k)\,\dir k. \,.
\end{eqnarray*}
For example, calculations for $N=4$ result in
\begin{eqnarray}\label{eq:symmats}
\overline{X}_F &=& \left[
         \begin{array}{ccccc}0&4&0&0&0\cr 
                  1&0&3&0&0\cr 
                  0&2&0&2&0\cr 
                  0&0&3&0&1\cr
                  0&0&0&4&0\cr\end{array}\right]\\
\overline{X}_G  &=& \left[\begin{array}{rrrrr}4&0&0&0&0\cr
                     0&2&0&0&0\cr
                     0&0&0&0&0\cr
                     0&0&0&-2&0\cr
                     0&0&0&0&-4\cr\end{array}\right] \\
\overline{H}_4 &=& \left[\begin{array}{rrrrr}
                    1&4&6&4&1\cr
                    1&2&0&-2&-1\cr 
                    1&0&-2&0&1\cr
                    1&-2&0&2&-1\cr
                    1&-4&6&-4&1\cr \end{array} \right]
 \label{eq:symm2}
\end{eqnarray}
\bigskip

Since $\overline{X}_G$ is the result of diagonalizing
$\overline{X}_F$, we observe that\bigskip

\begin{corollary}
The spectrum of $\overline{X}_F$ is
$N,N-2,\ldots,2-N,-N$, coinciding with the support 
of the classical random walk.
\end{corollary}
~
\\
\\

\noindent
{\bf Remark on the shuffling map. }
Notice that the top row of $(I+tF)^{\otimes N}$ is exactly $t^{w(k)}$, where
$w(k)$ is the binary shuffling function of section \S\ref{sec:result}. 
Each time one tensors with $I+tF$, 
the original top row is reproduced, then  concatenated with a replica 
of itself modified in that each entry picks up a factor of $t$ 
(compare with equation (\ref{eq:dynamic})).
And, collapsing to the symmetric tensor space,
the top row will have entries ${N\choose k}t^k$. This
follows as well by direct calculation of the $0^{\mathrm{th}}$ component
matrix elements in degree $N$, namely by expanding $(x+ty)^N$.\\

We continue with some areas where Krawtchouk polynomials/matrices play a 
r\^ole, very often not explicitly recognized in the original contexts.

%%%%%%%%%%%%%%% Intro to the world of Kravchukiana %%%%%%%%%%%%%%%

\section{Ehrenfest urn model}

In order to explain how the apparent irreversibility of the second law 
of thermodynamics arises from reversible statistical physics, 
the Ehrenfests introduced a so-called urn model, variations of which have been 
considered by many authors (\cite{ Kac, KMG, Voi}).
\\

We have an urn with $N$ balls. Each ball can be in two states 
represented by, say, being lead or gold.  At each time $k\in\mathbb{N}$, 
a ball is drawn at random, changed by a Midas-like touch into the opposite state
(gold $\leftrightarrow$ lead) and placed back in the urn.
The question is of course about the distribution of states ---
and this leads  to Krawtchouk matrices.
\\

Represent the states of the model by vectors in $\mathbb{R}^{n+1}$, 
namely by the state of $k$ gold balls by
\begin{equation}\label{e:states}
\begin{array}{rrcl}
\mathbf{v}_k = [\;0 & 0 &  \cdots  \quad 1 \quad \cdots & 0 \;]^\top \\
  &   & \uparrow \\
  &   & \hbox to 25pt{} k^{\hbox{th}}\hbox{ position}
\end{array}
\end{equation}  
In the case of, say, $N=3$, we have 4 states
$$
\begin{array}{l} \hbox{\small \textsf{0 gold balls}}\\ 
                 \hbox{\small \textsf{3 lead balls}}  \end{array}
   = \left[ \begin{array}{c} 1 \\ 0 \\ 0 \\ 0  \end{array}\right]
\qquad
\begin{array}{l} \hbox{\small\textsf{1 gold ball}}\\ 
                 \hbox{\small\textsf{2 lead balls}} \end{array}
   = \left[ \begin{array}{c} 0 \\ 1 \\ 0 \\ 0 \end{array}\right]
\qquad
\ldots
\qquad
\begin{array}{l} \hbox{\small\textsf{3 gold balls}}\\ 
                 \hbox{\small\textsf{0 lead balls}}  \end{array}
   =  \left[ \begin{array}{c} 0 \\ 0 \\ 0 \\ 1  \end{array}\right]
$$
It is easy to see that the matrix of elementary state change in this case 
is
$$
  \left[\begin{array}{rrrr} 
                0 & \frac{1}{3} & 0 & 0 \cr
                1 & 0 & \frac{2}{3} & 0 \cr
                0 & \frac{2}{3} & 0 & 1 \cr
                0 & 0 & {1\over3} & 0 \cr \end{array}\right]
\quad = \quad 
\frac{1}{3}\,
  \left[\begin{array}{rrrr} 
                0 & 1 & 0 & 0 \cr
                3 & 0 & 2 & 0 \cr
                0 & 2 & 0 & 3 \cr
                0 & 0 & 1 & 0 \cr \end{array}\right]
\quad = \quad\frac{1}{3}\, A^{(3)} \ ,
$$ 
and in general, we have the {\bf Kac matrix} with off-diagonals in arithmetic progression 
$1,2,3,...$ descending and ascending, respectively:
$$
A^{(N)} =
%\frac{1}{N}
  \left[\begin{array}{ccccccc} 
    0 & 1       & 0   & 0 &\cdots& 0&0 \\
    N & 0       & 2   & 0 &\cdots& 0&0 \\
    0 & N-1     & 0   & 3 &\vdots& 0&0 \\
    0 & 0       & N-2 & 0 &\ddots& 0&0 \\ 
  \vdots&\vdots&\vdots&\ddots&\ddots&\ddots &0\\
    0 & 0       & 0   & 0 &\ddots&0& N \\ 
    0 & 0       & 0   & 0 &\cdots&1& 0  
\end{array}\right]
$$ 
It turns out that the spectral properties of the Kac matrix 
involve Krawtchouk matrices, namely, the {\it collective solution} 
to the eigenvalue problem $Av=\lambda v$ is   
$$
A^{(N)}K^{(N)} =  K^{(N)} \Lambda^{(N)}
$$
where
$\Lambda^{(N)}$ is the $(N+1)\times(N+1)$ diagonal matrix with entries
$\Lambda^{(N)}_{ii}=N-2i$ 
$$
\Lambda^{(N)} =
  \left[\begin{array}{ccccccc} 
      N &     &    &  &  & \\
     & N-2 &    & & (\ast)& \\
     &     & N-4&   &&\\
     &     &    & \ddots &&\\ 
     &    (\ast)  &   &      &2-N  &\\ 
     &     &    &     &    & -N  
\end{array}\right]
$$ 
the  $(\ast)$'s denoting blocks of zeros.\\

To illustrate, for $N=3$ we have
$$
  \left[\begin{array}{rrrr} 
                0 & 1 & 0 & 0 \cr
                3 & 0 & 2 & 0 \cr
                0 & 2 & 0 & 3 \cr
                0 & 0 & 1 & 0 \cr \end{array}\right]
  \left[\begin{array}{rrrr} 
     1 &  1 &  1 &  1 \cr
     3 &  1 & -1 & -3 \cr
     3 & -1 & -1 &  3 \cr
     1 & -1 &  1 & -1 \cr \end{array}\right]
=
  \left[\begin{array}{rrrr} 
     1 &  1 &  1 &  1 \cr
     3 &  1 & -1 & -3 \cr
     3 & -1 & -1 &  3 \cr
     1 & -1 &  1 & -1 \cr \end{array}\right]
  \left[\begin{array}{rrrr} 
          3 & 0 & 0 & 0 \cr
          0 & 1 & 0 & 0 \cr
          0 & 0 & -1 & 0 \cr
          0 & 0 & 0 & -3 \cr \end{array}\right]
$$

To see this in general, we note that, cf. equations (\ref{eq:symmats}--\ref{eq:symm2}), 
these are the same operators appearing in the quantum random walk model, namely, we discover that
$\Lambda^{(N)}=\overline{X}_G$, $A^{(N)}=\overline{X}_F^\top$. Now, recalling
$K^{(N)}=\overline{H}_N^\top$, taking transposes in equation (\ref{eq:spec}) yields
$$
A^{(N)}\, K^{(N)} = K^{(N)}\, \Lambda^{(N)} \qquad\hbox{and}\qquad K^{(N)}\, A^{(N)} = \Lambda^{(N)}\, K^{(N)} 
$$
which is the spectral analysis of $A^{(N)}$ from both the left and the right.
Thus, e.g.,  the columns of the Krawtchouk matrix are eigenvectors of the Ehrenfest model
with $N$ balls  
where the $k^{\hbox{th}}$ column $\?v?_k:= (K_{\cdot\,k})$ has corresponding eigenvalue
$\lambda_k=(N-2k)/N$.  
\bigskip

\noindent
{\bf Remarks }
\begin{enumerate}
\item Clearly, the Ehrenfest urn problem can be expressed in other
terms. For instance, it can be reformulated as a
random walk on an $N$-dimensional cube.
Suppose an ant walks on the cube, 
choosing at random an edge to progress to the next vertex.  
Represent the states by vectors in $Z=\mathbb{Z}_2\times\cdots\times\mathbb{Z}_2$, $N$ factors.      
The equivalence of the two problems comes via the 
correspondence of states 
$$
Z\ni [\;a_1\;a_2\ldots a_N\;]
\longrightarrow \?v?_w\in \mathbb{R}^{N+1}
$$
where $w = \sum a_i$ is the weight of the vector calculated in $\mathbb{N}$,
see (\ref{e:states}).\\

\item  The urn model in the appropriate limit as $N\to\infty$ leads to a diffusion model on the line,
the discrete distributions converging to the diffusion densities. See Kac' article (\cite{Kac}).\\

\item There is a rather unexpected connection of the urn model 
with finite-dimensional representations of the Lie algebra $sl(2)\cong so(2,1)$.
Indeed, introduce a new matrix by the commutator:
$$
\overline{A} = \frac{1}{2}\;[A,\Lambda]
$$
The matrix $\overline{A}$ is a skew-symmetric version of $A$.
For $N=3$, it is 
$$
\overline{A}= \left[\begin{array}{rrrr} 
                0 & -1 & 0 & 0 \cr
                3 & 0 & -2 & 0 \cr
                0 & 2 & 0 & -3 \cr
                0 & 0 & 1 & 0 \cr \end{array}\right]
$$
It turns out that the triple $A$, $\overline{A}$ and $\Lambda$
is closed under commutation, thus forms a Lie algebra, namely
$$
\hbox{span}\;\{\;A,\;\overline{A},\;\Lambda\;\} 
             \cong so(2,1)
             \cong sl(2,\mathbb{R})
$$
with commutation relations
$$
[ A, \overline{A} ]     = 2\Lambda \,,\qquad
[\overline{A},\Lambda ] = 2A \,,\qquad
[ \Lambda, A ]          = -2 \overline{A} 
$$  
\end{enumerate}

	%%%%%%%%%%%%%% Krawtchouk matrices and classical random walk %%%%%%%%%%%%%%%%

\section{Krawtchouk matrices and classical random walks}
\label{sec:classical}

In this section we will give a probabilistic meaning to
the Krawtchouk matrices and illustrate some connections with classical random walks.\\

\subsection{ Bernoulli random walk }
Let $X_i$ be independent symmetric Bernoulli random variables taking
values $\pm1$. Let $x_N=X_1+\cdots+X_N$ be the associated random
walk starting from $0$. Now observe that the generating function of
the elementary symmetric functions in the $X_i$ is a martingale, in
fact a discrete exponential martingale:
$$
   M_N = \prod_{i=1}^N(1+vX_i)=\sum_k v^k 
           a_k(X_1,\ldots,X_N)
$$
where $a_k$ denotes the $k^{\mathrm{th}}$ elementary symmetric
function. 
The martingale property is immediate since each $X_i$ has mean $0$.
Refining the notation by setting $a_k^{(N)}$ to denote the $k^{\mathrm{th}}$ elementary symmetric
function in the variables $X_1,\ldots, X_N$, multiplying $M_N$ by $1+vX_{N+1}$ yields the recurrence

$$
a_k^{(N+1)} = a_k^{(N)}+a_{k-1}^{(N)}\,X_{N+1}
$$
which, with the boundary conditions $a_k^{(0)}=0$, for $k>0$, $a_0^{(n)}=1$ for all $n\ge0$,  yields, for $k>0$,
$$
a_k^{(N+1)} = \sum_{j=0}^N a_{k-1}^{(j)}\,X_{j+1}
$$
that is, these are discrete or {\sl prototypical iterated stochastic integrals} and thus the simplest
example of Wiener's homogeneous chaoses. \\

Suppose that at time $N$, the number of the $X_i$ that are equal to $-1$
is $j_N$, with the rest equal to $+1$. Then  $j_N= (N-x_N)/2$ 
and $M_N$ can be expressed solely in terms of $N$ and $x_N$, or, 
equivalently, of $N$ and $j_N$
$$
       M_N = (1+v)^{N-j_N}(1-v)^{j_N} 
           = (1+v)^{(N+x_N)/2}(1-v)^{(N-x_N)/2}  
$$

From the generating function for the Krawtchouk
matrices, equation (\ref{eq:genkraw}), follows 
$$
    M_N = \sum_iv^iK^{(N)}_{i,j_N}
$$
so that as functions on the Bernoulli space, each sequence of random
variables $K^{(N)}_{i,j_N}$ is a martingale. \\

Now we can derive two basic recurrences. From a given
column of $K^{(N)}$, to get the corresponding column in 
$K^{(N+1)}$,  we have the Pascal's triangle recurrence:
$$
  \FFF N{i-1\;j} + \FFF N{i  \;j}  =  \FFF {N+1}{i\;j}
$$
This follows in the probabilistic setting by writing
$M_{N+1}=(1+vX_N)M_N$ and remarking that for $j$ to remain constant,
$X_N$ must take the value $+1$. The martingale property is more interesting in the
present context. We have
$$
\FFF N{i\,j_N}= E(\FFF N{i\,j_{N+1}}|X_1,\ldots,X_N)
             =\frac12\,\left(\FFF{N+1}{i\,j_N+1}+\FFF{N+1}{i\,j_N}\right)
$$
since half the time $X_{N+1}$ is $-1$, increasing $j_N$ by 1, and
half the time $j_N$ is unchanged. Thus, writing $j$ for $j_N$,
$$
      \FFF N{ij}=\frac12\,\left(\FFF{N+1}{i\,j+1}+\FFF{N+1}{ij}\right)
$$
which may be considered as a `reverse Pascal'.
\bigskip
\subsubsection{Orthogonality}
As noted above --- here with a slightly simplified notation ---
it is natural to use variables $(x,N)$, with $x$ denoting the
position of the random walk after $N$ steps. Writing $K_{\alpha}(x,N)$ for the
Krawtchouk \pl s in these variables, cf. equation (\ref{eq:poly}), we have  the generating function
$$G(v)=\sum_{\alpha=0}^N \.v.\alpha.K_{\alpha} (x,N) =\.(1+v).(N+x)/2.
             \.(1-v).(N-x)/2.$$
The expansion
\begin{equation}\label{eq:hypergeo}
\.(1-v).y-a.\.(1-(1-R)v).-y.=\sum_{n=0}^\infty
         \ov v^n.n!.\,(a)_n\,\tfo -n.y.a.R. 
\end{equation}
with $(a)_n=\Gamma(a+n)/\Gamma(a)$, yields the identification as hypergeometric functions

$$
K_{\alpha}(x,N)=\kck N.\alpha.\,\tfo-\alpha.(x-N)/2.-N.2.
$$
The calculation
$$\ln G(v)\,G(w)\rn=\prod\ln1+(v+w)X_j+vwX_j^2\rn
                    =(1+vw)^N$$
exhibits the orthogonality of the $K_\alpha$  
if one observes that after taking expectations only terms in the product $vw$ remain.  Thus, the $K_\alpha$ are
notable for two important features:\\
\begin{enumerate}
\item They are the iterated integrals (sums) of the Bernoulli
process.
\item They are orthogonal \pl s with respect to the binomial
distribution.
\end{enumerate}

\subsection{Multivariate Krawtchouk polynomials}

The probabilistic approach may be carried out for general finite
probability spaces.
Fix an integer $d>0$ and $d$ values $\{\xi_0,\ldots,\xi_\delta\}$, with the convention $\delta=d-1$.
Take a sequence of independent identically distributed random variables having distribution
$P(X=\xi_j)=p_j,\,0\le j\le\delta$. Denote the mean and variance of the $X_i$ by $\mu$ and 
$\sigma^2$ as usual.\\
 
For $N>0$, we have the martingale
$$M_N=\prod_{j=1}^N (1+v(X_j-\mu))$$
We now switch to the multiplicities as variables. Set
$$n_j=\sum_{k=1}^N \?1?_{\{X_k=\xi_j\}}$$
the number of times the value $\xi_j$ is taken. Thus the generating function
$$G(v)=\prod_{j=0}^\delta(1+v(\xi_j-\mu))^{n_j}=\sum_{\alpha=0}^N
\.v.\alpha.K_{\alpha}(n_0,\ldots,n_\delta)$$
defines our generalized Krawtchouk \pl s. One quickly gets
 
\begin{proposition}Denoting the multi-index $\?n?=(n_0,\ldots,n_\delta)$
 and by $\?e_j?$ the standard basis on $\mathbb{Z}^d$, Krawtchouk \pl s satisfy the
recurrence
$$K_{\alpha}(\?n?+\?e_j?)=K_{\alpha}(\?n?)+(\xjm)K_{\alpha-1}(\?n?)$$
\end{proposition}
We also find by binomial expansion
\begin{proposition}
$$K_{\alpha}(n_0,\ldots,n_\delta)=\sum_{\vert\?k?\vert=\alpha}\prod_j\kck n_j.k_j.
           \.(\xjm).k_j.$$
where $\vert\?k?\vert=\sum\limits_{j=0}^\delta k_j$.
\end{proposition}
 
There is an interesting connection with the multivariate hypergeometric functions of Appell and Lauricella.
The Lauricella \pl s $F_B$ are defined by
$$\tfb-\?r?.\?b?.t.\?s?.=\sum_{\?k?\in\mathbb{N}^\delta}
{(-\?r?)_\?k?(\?b?)_\?k?\over (t)_{\vert\?k?\vert}\?k?!}\,\.\?s?.\?k?.
  $$
with, e.g., $\?r?=(r_1,\ldots,r_\delta),\,(\?r?)_{\?k?}=
 (r_1)_{k_1}(r_2)_{k_2}\cdots(r_\delta)_{k_{\delta}}$
  for multi-index $\?k?$,
also $\.\?s?.\?k?.=\.s_1.k_1.\cdots\.s_\delta.k_\delta.,\,{\rm and}\,
\?k?!=k_1!\cdots k_\delta!\,$. Note that $t$ is a single variable. The generating function
of interest here is
\begin{equation} \label{eq:laur}
\.(1-\sum v_i).\sum b_j-t.\,\prod_j\.(1-\sum v_i+s_jv_j).-b_j.
=\sum_{\?r?\in\mathbb{N}^\delta} {\.\?v?.\?r?.(t)_{\vert\?r?\vert}\over \?r?!}\,
 \tfb-\?r?.\?b?.t.\?s?.
\end{equation}
a multivariate version of (\ref{eq:hypergeo}).
\begin{proposition}{Let $N=|\?n?|$. If $\xi_0=0$, then, 
$$K_{\alpha}(\?n?)=(-N)_\alpha\,\sum_{\vert\?r?\vert=\alpha}{\prod\.(p_j\xi_j).r_j.
\over \?r?!}\,\tfb-\?r?.-\?n?.-N.{{1\over p_1},\ldots,{1\over p_\delta}}.
$$}
\end{proposition}
{\sl Proof } Let $v_j=vp_j\xi_j,\,b_j=-n_j,\,t=-N,\,s_j=p_j^{-1}$ in (\ref{eq:laur}),
for $1\le j\le\delta$. Note that $\sum v_j=v\mu$, $\sum b_j-t=
N-(\sum\limits_{1\le j\le \delta} n_j)=n_0$.\qed
\bigskip
 
Orthogonality follows similar to the binomial case:
\begin{proposition}The Krawtchouk \pl s $K_{\alpha}(n_0,\ldots,n_\delta)$ are orthogonal
with respect to the induced multinomial distribution. In fact, \\ with $N=|\?n?|$,
$$\ln K_{\alpha}K_{\beta}\rn=\delta_{\alpha\beta}\,\.\sigma.2\alpha.\,\kck N.\alpha.$$
\end{proposition}
{\sl Proof }
\begin{eqnarray*}
\ln G(v)\,G(w)\rn&=&\sum\kck N.{n_0,\ldots,n_\delta}.\.p_0.n_0.
\cdots\.p_{\delta}.n_{\delta}.\,\prod\left(1+(v+w)(\xjm)+vw(\xjm)^2
\right)^{n_j}\cr
&=&\left(\sum\left(p_j+(v+w)p_j(\xjm)+vwp_j(\xjm)^2\right)\right)^N\cr
\end{eqnarray*}

Thus, $\ln G(v)\,G(w)\rn=(1+vw\.\sigma.2.)^N$. This shows orthogonality
and yields the squared norms as well.\qed
\bigskip

%%%%%%%%%%%%% Kravchukiana %%%%%%%%%%%%%
\section{``Kravchukiana'' or the World of Krawtchouk Polynomials}
About the year 1995, we held a seminar on Krawtchouk polynomials at Southern Illinois University.
As we continued, we found more and more properties and connections with various areas of mathematics.\\

Eventually, by the year 2000 the theory of quantum computing had been developing with serious
interest in the possibility of implementation, at the present time of MUCH interest. Sure enough,
right in the middle of everything there are our flip operators, su(2), etc., etc.  --- same ingredients
making up the Krawtchouk universe. Well, we can only report that how this all fits together is still quite open.
Of special note is the idea of a hardware implementation of a Krawtchouk transform.
A beginning in this direction may be found in the just-published article with Schott, Botros, and
Yang \cite{BY}. \\

At any rate, for the present we list below the topics which are central to our program.
They are the basis of the  {\bf Krawtchouk Encyclopedia}, still in development;
 we are in the process of filling in the blanks.
 An extensive web resource for Krawtchouk polynomials we recommend is Zelenkov's site:\\

{\tt http://www.geocities.com/orthpol/}\\

Note that we do not mention work in areas less familiar to us, 
notably that relating to $q$-Krawtchouk polynomials, such as in \cite{St}.\\

{\it We welcome contributions. If you wish either to send a reference to your paper(s)
on Krawtchouk polynomials or contribute an article, please contact one of us ! }\\

Our email: {\tt pfeinsil@math.siu.edu} or {\tt jkocik@math.siu.edu}.\\

\subsection{Krawtchouk Encyclopedia}

Here is a list of topics currently in the Krawtchouk Encyclopedia.\\

\begin{enumerate}

\item Pascal's Triangle

\item Random Walks  
\begin{itemize} 
\item  Path integrals
\item  A, K, and $\Lambda$
\item  Nonsymmetric Walks
\item  Symmetric Krawtchouk matrices and binomial expectations 
 \end{itemize}

\item Urn Model  

\begin{itemize} 
\item   Markov chains 
\item  Initial and invariant distributions 
\end{itemize}

\item Symmetric Functions. Energy  

\begin{itemize} 
\item   Elementary symmetric functions and determinants 
\item  Traces on Grassman algebras

 \end{itemize}

\item Martingales

\begin{itemize} 
\item   Iterated integrals 
\item  Orthogonal functionals 
\item  Krawtchouk polynomials and multinomial distribution 
 \end{itemize}

\item Lie algebras and Krawtchouk polynomials

\begin{itemize}
\item   so(2,1) explained 
\item  so(2,1) spinors 
\item  Quaternions and Clifford algebras 
\item  S and so(2,1) tensors 
\item Three-dimensional simple Lie algebras 
 \end{itemize}

\item Lie Groups. Reflections

\begin{itemize}
\item  Reflections 
\item  Krawtchouk matrices as group elements 
 \end{itemize}

\item Representations

\begin{itemize}
\item   Splitting formula 
\item  Hilbert space structure 
 \end{itemize}

\item Quantum Probability and Tensor Algebra

\begin{itemize}
\item Flip operator and quantum random walk 
\item  Krawtchouk matrices as eigenvectors 
\item  Trace formulas. MacMahon's Theorem  
\item  Chebyshev polynomials  
 \end{itemize}

\item Heisenberg Algebra

\begin{itemize}
\item  Representations of the Heisenberg algebra 
\item  Raising and velocity operator. Number operator 
\item Evolution structure. Hamiltonian. 
\item Time-zero polynomials 
 \end{itemize}

\item Central Limit Theorem

\begin{itemize}
\item  Hermite polynomials 
\item  Discrete stochastic differential equations 
 \end{itemize}

\item Clebsch-Gordan Coefficients
\begin{itemize}
\item   Clebsch-Gordan coefficients and Krawtchouk polynomials 
\item  Racah coefficients 
 \end{itemize}

\item Orthogonal Polynomials   

\begin{itemize}
\item Three-term recurrence in terms of A, K, Lambda 
\item Nonsymmetric case 
 \end{itemize}

\item Krawtchouk Transforms

\begin{itemize}
\item  Orthogonal transformation associated to K 
\item  Exponential function in Krawtchouk basis 
\item Krawtchouk transform
 \end{itemize}

\item Hypergeometric Functions

\begin{itemize}
\item   Krawtchouk polynomials as hypergeometric functions 
\item  Addition formulas 
 \end{itemize}

\item Symmetric Krawtchouk Matrices

\begin{itemize}
\item The matrix T 
\item  S-squared and trace formulas 
\item  Spectrum of S 
 \end{itemize}

\item Gaussian Quadrature

\begin{itemize}
\item  Zeros of Krawtchouk polynomials 
\item  Gaussian-Krawtchouk summation 
 \end{itemize}

\item Coding Theory

\begin{itemize}
\item  MacWilliams' theorem 
\item Association schemes 
 \end{itemize}

\item Appendices  

\begin{itemize}
\item  K and S matrices for N from 1 to 14 
\item  Krawtchouk polynomials in the variables x,N/i,j/j,N for N from 1 to 20 
\item  Eigenvalues of S 
\item  Remarks on the multivariate case 
\item  Time-zero polynomials 
\item  Mikhail Philippovitch Krawtchouk: a biographical sketch 
\end{itemize}

\end{enumerate}

\pagestyle{empty}
		%%%%%%%%%%%%%%%%%%%%%%%%%%%%%%%%%%%
		%%%%%%%%%%%%% krav matrices %%%%%%%%%%%%%
		%%%%%%%%%%%%%%%%%%%%%%%%%%%%%%%%%%%%
%==================  krav matrices ==================================
\section{Appendix}
\subsection{Krawtchouk matrices}

\begin{eqnarray*}
\KM^{(0)}&=&\left[\begin{array}{r} 1 \end{array}\right]   
\\
\\
\KM^{(1)}&=&\left[\begin{array}{rr} 1 &  1 \cr
                           1 & -1 \cr\end{array}\right]    
\\
\\
\KM^{(2)}&=& \left[\begin{array}{rrr}  1 &  1 &  1 \cr
                             2 &  0 & -2 \cr
                             1 & -1 &  1 \cr \end{array}\right]
\\
\\
\KM^{(3)}&=&
       \left[\begin{array}{rrrr}
                      1 &  1 &  1  &  1 \cr
                      3 &  1 & -1  & -3 \cr
                      3 & -1 & -1  &  3 \cr
                      1 & -1 &  1  & -1 \cr 
             \end{array}\right]
\\
\\
\KM^{(4)}&=&
       \left[\begin{array}{rrrrr} 1 &  1 &  1  &  1  &  1 \cr
                      4 &  2 &  0  & -2  & -4 \cr
                      6 &  0 & -2  &  0  &  6 \cr
                      4 & -2 &  0  &  2  & -4 \cr
                      1 & -1 &  1  & -1  &  1 \cr \end{array}\right]
\\
\\
\KM^{(5)}&=&
       \left[\begin{array}{rrrrrr} 1  &  1 &  1  &  1  &  1 &   1 \cr
                      5  &  3 &  1  & -1  & -3 &  -5 \cr
                     10  &  2 & -2  & -2  &  2 &  10 \cr
                     10  & -2 & -2  &  2  &  2 & -10 \cr
                      5  & -3 &  1  &  1  & -3 &   5 \cr
                      1  & -1 &  1  & -1  &  1 &  -1 \cr \end{array}\right]
\\
\\
\KM^{(6)}&=&
       \left[\begin{array}{rrrrrrr} 
                      1  &  1 &  1  &  1  &  1 &  1  &   1 \cr
                      6  &  4 &  2  &  0  & -2 & -4  &  -6 \cr
                     15  &  5 & -1  & -3  & -1 &  5  &  15 \cr
                     20  &  0 & -4  &  0  &  4 &  0  & -20 \cr
                     15  & -5 & -1  &  3  & -1 & -5  &  15 \cr
                      6  & -4 &  2  &  0  & -2 &  4  &  -6 \cr
                      1  & -1 &  1  & -1  &  1 & -1  &   1 \cr
\end{array}\right] 
\end{eqnarray*}

\centerline{{\bf Table 1}}

%=============   symmetric krav matrices =============================
\newpage
\subsection{Symmetric Krawtchouk matrices}
\begin{eqnarray*}
S^{(0)}&=&\left[\begin{array}{r} 1 \end{array}\right]
\\
\\
S^{(1)}&=&\left[\begin{array}{rr} 1 &  1 \cr
                        1 & -1 \cr\end{array}\right]
\\
\\
S^{(2)}&=& \left[\begin{array}{rrr}  1 &  2 &  1 \cr
                          2 &  0 & -2 \cr
                          1 & -2 &  1 \cr \end{array}\right]
\\
\\
S^{(3)}&=&
      \left[\begin{array}{rrrr} 1 &  3 &  3  &  1 \cr
                     3 &  3 & -3  & -3 \cr
                     3 & -3 & -3  &  3 \cr
                     1 & -3 &  3  & -1 \cr \end{array}\right]
\\
\\
S^{(4)}&=&
      \left[\begin{array}{rrrrr} 1 &  4 &  6  &  4  &  1 \cr
                     4 &  8 &  0  & -8  & -4 \cr
                     6 &  0 &-12  &  0  &  6 \cr
                     4 & -8 &  0  &  8  & -4 \cr
                     1 & -4 &  6  & -4  &  1 \cr \end{array}\right]
\\
\\
S^{(5)}&=&
      \left[\begin{array}{rrrrrr} 1  &  5 & 10 &  10  &   5 &   1 \cr
                     5  & 15 & 10 & -10  & -15 &  -5 \cr
                    10  & 10 &-20 & -20  &  10 &  10 \cr
                    10  &-10 &-20 &  20  &  10 & -10 \cr
                     5  &-15 & 10 &  10  & -15 &   5 \cr
                     1  & -5 & 10 & -10  &   5 &  -1 \cr \end{array}\right]
\\
\\
S^{(6)}&=&
       \left[\begin{array}{rrrrrrr} 1  &   6 &  15  &  20 &  15 &   6 &   1
\cr
                      6  &  24 &  30  &   0 & -30 & -24 &  -6 \cr
                     15  &  30 & -15  & -60 & -15 &  30 &  15 \cr
                     20  &   0 & -60  &   0 &  60 &   0 & -20 \cr
                     15  & -30 & -15  &  60 & -15 & -30 &  15 \cr
                      6  & -24 &  30  &   0 & -30 &  24 &  -6 \cr
                      1  &  -6 &  15  & -20 &  15 &  -6 &   1 \cr
\end{array}\right]
\end{eqnarray*}

\centerline{{\bf Table 2} }

%=============   hadamard matrices =============================
\newpage
\def\b{\circ}   \def\a{\bullet}

\subsection{Sylvester-Hadamard matrices}

\begin{eqnarray*}
H^{(0)}&=&\left[\begin{array}{c}\a \cr\end{array}\right]
\\
H^{(1)}&=&
\left[\begin{array}{cc}
      \a &\a  \cr
      \a &\b  \cr
\end{array}\right]
\\
H^{(3)}&=&
\left[\begin{array}{cccc}
\a &\a &\a &\a  \cr
\a &\b &\a &\b   \cr 
\a &\a &\b &\b   \cr
\a &\b &\b &\a   \cr
\end{array}\right]
\\
\\
H^{(4)}&=&
\left[\begin{array}{cccccccc}
\a &\a &\a &\a  &\a &\a &\a &\a    \cr
\a &\b &\a &\b  &\a &\b &\a &\b    \cr
\a &\a &\b &\b  &\a &\a &\b &\b    \cr
\a &\b &\b &\a  &\a &\b &\b &\a    \cr
\a &\a &\a &\a  &\b &\b &\b &\b    \cr 
\a &\b &\a &\b  &\b &\a &\b &\a    \cr 
\a &\a &\b &\b  &\b &\b &\a &\a    \cr 
\a &\b &\b &\a  &\b &\a &\a &\b    \cr 
\end{array}\right] 
\\
\\
H^{(5)}&=&
\left[\begin{array}{cccccccccccccccc}
\a &\a &\a &\a  &\a &\a &\a &\a  &\a &\a &\a &\a  &\a &\a &\a &\a  \cr
\a &\b &\a &\b  &\a &\b &\a &\b  &\a &\b &\a &\b  &\a &\b &\a &\b  \cr 
\a &\a &\b &\b  &\a &\a &\b &\b  &\a &\a &\b &\b  &\a &\a &\b &\b  \cr
\a &\b &\b &\a  &\a &\b &\b &\a  &\a &\b &\b &\a  &\a &\b &\b &\a  \cr
\a &\a &\a &\a  &\b &\b &\b &\b  &\a &\a &\a &\a  &\b &\b &\b &\b  \cr
\a &\b &\a &\b  &\b &\a &\b &\a  &\a &\b &\a &\b  &\b &\a &\b &\a  \cr
\a &\a &\b &\b  &\b &\b &\a &\a  &\a &\a &\b &\b  &\b &\b &\a &\a  \cr
\a &\b &\b &\a  &\b &\a &\a &\b  &\a &\b &\b &\a  &\b &\a &\a &\b  \cr
\a &\a &\a &\a  &\a &\a &\a &\a  &\b &\b &\b &\b  &\b &\b &\b &\b  \cr
\a &\b &\a &\b  &\a &\b &\a &\b  &\b &\a &\b &\a  &\b &\a &\b &\a  \cr
\a &\a &\b &\b  &\a &\a &\b &\b  &\b &\b &\a &\a  &\b &\b &\a &\a  \cr
\a &\b &\b &\a  &\a &\b &\b &\a  &\b &\a &\a &\b  &\b &\a &\a &\b  \cr
\a &\a &\a &\a  &\b &\b &\b &\b  &\b &\b &\b &\b  &\a &\a &\a &\a  \cr
\a &\b &\a &\b  &\b &\a &\b &\a  &\b &\a &\b &\a  &\a &\b &\a &\b  \cr
\a &\a &\b &\b  &\b &\b &\a &\a  &\b &\b &\a &\a  &\a &\a &\b &\b  \cr
\a &\b &\b &\a  &\b &\a &\a &\b  &\b &\a &\a &\b  &\a &\b &\b &\a  \cr
\end{array}\right]
\end{eqnarray*}

\centerline{{\bf Table 3} }
\bigskip
\hfill{\footnotesize
Replace $\bullet$ with 1 and $\circ$ with $-1$ to obtain Sylvester-Hadamard matrices.}\hfill
%\newpage

%============================================
	%%%%%%%%%%%%% thebibliography %%%%%%%%%%%%%
%============================================

\end{document}